\documentclass{svjour3}
\usepackage{graphicx}
\usepackage{eurosym}
\usepackage{amsmath}
\usepackage{amsfonts}
\usepackage{amssymb}
\usepackage{hyperref}
\usepackage{setspace}
\usepackage{mathpazo}
\usepackage{stmaryrd}
\usepackage{latexsym}

\RequirePackage{fix-cm}
\smartqed
\begin{document}

\title{
\date{Received: date / Accepted: date}

$m$-adic residue codes over $\mathbb{F}_q[v]/(v^s-v)$ and their
applications to quantum codes}

\author{Ferhat K\"{u}r\"{u}z \and Mustafa Sar{\i} \and Mehmet E. K\"{o}ro\u{g}lu}

\titlerunning{$m$-adic residue codes over a class of nonchain rings and their applications}
\authorrunning{F. K\"{u}r\"{u}z \and M. Sar{\i} \and M. E. K\"{o}ro\u{g}lu}
\institute{Ferhat K\"{u}r\"{u}z \at
              Department of Computer Engineering, Gelisim University, Avcılar 34310, Istanbul-Turkey\\
              \email{fkuruz@gelisim.edu.tr} \\
           Mustafa Sar{\i} \at
              Department of Mathematics, Yildiz Technical University, Esenler 34220, Istanbul-Turkey \\
              \email{musari@yildiz.edu.tr}\\
           Mehmet E. K\"{o}ro\u{g}lu \at
              Department of Mathematics, Yildiz Technical University, Esenler 34220, Istanbul-Turkey \\
              \email{mkoroglu@yildiz.edu.tr}}

\date{Received: date / Accepted: date}
\maketitle

\begin{abstract}
Due to their rich algebraic structure, cyclic codes have a great deal of
significance amongst linear codes. Duadic codes are the generalization of
the quadratic residue codes, a special case of cyclic codes. The $m$-adic
residue codes are the generalization of the duadic codes. The aim of this
paper is to study the structure of the $m$-adic residue codes over the
quotient ring $\frac{{{\mathbb{F}_q}\left[ v \right]}}{{\left\langle {{v^s} - v} \right\rangle }}$. We determine the idempotent
generators of the $m$-adic residue codes over $\frac{{{\mathbb{F}_q}\left[ v \right]}}{{\left\langle {{v^s} - v} \right\rangle }}$.
We obtain some parameters of optimal $m$-adic residue codes over $\frac{{{\mathbb{F}_q}\left[ v \right]}}{{\left\langle {{v^s} - v} \right\rangle }}$ with respect to Griesmer bound for rings. Furthermore, we
derive a condition for $m$-adic residue codes over $\frac{{{\mathbb{F}_q}\left[ v \right]}}{{\left\langle {{v^s} - v} \right\rangle }}$ to contain their dual. By making use of a preserving-orthogonality Gray map defined in \cite{goyal}, we construct a
family of quantum error correcting codes from the Gray images of
dual-containing $m$-adic residue codes over $\frac{{{\mathbb{F}_q}\left[ v \right]}}{{\left\langle {{v^s} - v} \right\rangle }}$ and
give some examples to illustrate our findings.

\end{abstract}

\section{Introduction}

Because of their strong algebraic properties, the class of cyclic codes have
a great deal of significance amongst linear codes. As a special case of
cyclic codes, the class of quadratic residue codes plays an important role in
cyclic codes for theoretical and practical reasons and they have been
studied over finite fields for many years. The class of duadic codes is the
generalization of the quadratic residue codes. Recently, many researchers
have studied the structure of generalization of quadratic residue codes over
some special rings (see \cite{A4,A6,A3,A2,A5} for instance). The $m$-adic
residue codes are the generalization of the duadic codes (\cite{A1}). The $m$%
-adic residue codes are investigated and are found to have many of the
strong properties of the quadratic residue codes. For the first time in
literature, Brualdi and Pless generalized the quadratic residue codes, which
they called polyadic codes, by defining the $m$-adic residue codes (see \cite%
{A4}). Pless studied the idempotent generators of the polyadic codes in \cite%
{A2}. In \cite{A1}, by virtue of generator polynomials over finite fields,
Job defined $m$-adic residue codes. In \cite{Ling}, Ling and Xing
extended the definition of polyadic cyclic codes to include noncyclic
abelian codes, and obtained necessary and sufficient conditions for the
existence of non-degenerate polyadic codes. Chen et al. studied the
structure of polyadic constacyclic codes and gave a necessary and sufficient
condition for the existence of Type-I $m$-adic $\lambda$-constacyclic codes
\cite{Chen}. In \cite{G1}, Goyal and Raka studied the structure of $%
(1-2u^{3})$-constacyclic codes over the quotient ring $\frac{{{\mathbb{F}_p}\left[ u \right]}}{{\left\langle {{u^4} - u} \right\rangle }}$. Then, as a continuation of this paper (\cite{G1}) they
extended their results to the ring $\frac{{{\mathbb{F}_p}\left[ u \right]}}{{\left\langle {{u^m} - u} \right\rangle }}$ and
studied quadratic residue codes and their extensions \cite{goyalquadratic}.
Kuruz et al. studied the structure of $m$-adic residue codes over the
non-chain ring $\frac{{{\mathbb{F}_q}\left[ v \right]}}{{\left\langle {{v^2} - v} \right\rangle }}$ and investigated the relation between the generators of these
codes and the reversible DNA codes \cite{kuruz}.

On the other hand, there have been many studies on constructing quantum
error correcting codes from the codes over the rings, some of which are \cite{ashraf,bag,kai,qian1,sari2,sari1,tang,xunru}. In \cite{kai}, Kai and Zhu
studied the linear and cyclic codes over the ring ${\mathbb{F}_{4}}+u{%
\mathbb{F}_{4}}$, where $u^{2}=0$, and determined the parameters of quantum
codes obtained by transforming the cyclic codes over ${\mathbb{F}_{4}}+u{%
\mathbb{F}_{4}}$ onto $\mathbb{F}_{4}$ via an orthogonality-preserving Gray
map. In \cite{xunru}, the authors defined a Gray map from $\mathbb{F}_{2}+u%
\mathbb{F}_{2}+u^{2}\mathbb{F}_{2}$ to $\mathbb{F}_{2}^{3}$ that preserves
the orthogonality, and derived a family of quantum codes from
dual-containing cyclic codes over this ring. As a generalization of the
study \cite{xunru}, in \cite{sari1}, Sar{\i } and Siap gave a construction
for orthogonality-preserving Gray map from the ring $\frac{{{\mathbb{F}_2}\left[ u \right]}}{{\left\langle {u^s} \right\rangle }}$, $%
u^{s}=0$ to $\mathbb{F}_{2}^{s}$, and obtained an existence-condition for
dual-containing cyclic codes over the ring $\frac{{{\mathbb{F}_2}\left[ u \right]}}{{\left\langle {u^s} \right\rangle }}$. They also gave the parameters of quantum codes obtained from the cyclic codes over the ring $\frac{{{\mathbb{F}_2}\left[ u \right]}}{{\left\langle {u^s} \right\rangle }}$. In \cite{qian1}, Qian studied the quantum codes
obtained from cyclic codes over the ring $\mathbb{F}_{2}+v\mathbb{F}_{2}$, $%
v^{2}=v$. In \cite{ashraf}, the authors considered quantum codes as the Gray
images of the cyclic codes over the ring $\mathbb{F}_{3}+v\mathbb{F}_{3}$, $%
v^{3}=v$. In \cite{sari2}, Sar{\i } and Siap extended the results obtained
in \cite{qian1} over a class of nonchain rings $\frac{{{\mathbb{F}_p}\left[ v \right]}}{{\left\langle {v^p-v} \right\rangle }}$. More recently, in \cite{bag}, Bag \emph{et al.} obtained new
nonbinary quantum codes from the codes over the ring $\frac{{{\mathbb{F}_p}\left[ u \right]}}{{\left\langle {u^3-u} \right\rangle }}$. In \cite{tang}, Tang \emph{et al.} utilized $\left(
1+u\right) $-constacyclic codes over the ring ${\mathbb{F}_{{2^{m}}}}+u{%
\mathbb{F}_{{2^{m}}}}$ to construct quantum codes.

In this work, inspired by the above works, as a generalization of the paper
\cite{kuruz}, we study the structure of the $m$-adic residue codes over the
quotient ring $\mathcal{R}_{q,s}=\frac{{{\mathbb{F}_q}\left[ v \right]}}{{\left\langle {{v^s} - v} \right\rangle }}$. First, we
determine the idempotent generators of the $m$-adic residue codes over $%
\mathcal{R}_{q,s}$ and give an algebraic characterization for these codes.
Then, we obtain some parameters of optimal $m$-adic residue codes over $%
\mathcal{R}_{q,s}$ with respect to Griesmer bound for rings. Furthermore,
via CSS construction for quantum error correcting codes and the
orthogonality-preserving Gray map defined in \cite{goyal}, we apply $m$-adic
residue codes over $\mathcal{R}_{q,s}$ to obtain quantum codes and
illustrate the finding by giving some examples.

The rest of this paper is organized as follows. In Section 2, we present
some definitions and basic results of linear codes, cyclic codes and $m$%
-adic residue codes. In Section 3, we study the idempotent generators of $m$%
-adic residue codes over $\mathcal{R}_{q,s}$ and characterize these codes by
determining their algebraic structures. Further, we obtain some parameters
of optimal $m$-adic residue codes over $\mathcal{R}_{q,s},$ with respect to
Griesmer bound for rings. In Section 4, we determine the duals of $m$-adic
residue codes over $\mathcal{R}_{q,s}$ and give a condition for these codes
to contain their dual. By carrying dual-containing $m$-adic residue codes
over $\mathcal{R}_{q,s}$ via the Gray map given in \cite{goyal} and using
CSS code construction, we derive a family of quantum codes. Moreover, we
illustrate our finding by giving some examples. Section 5 is reserved for
conclusion and future remarks.

\section{Preliminaries}

In this section, we give some basic definitions and results about linear
codes, cyclic codes and $m$-adic residue codes and we introduce the
structure of the ring $\mathcal{R}_{q,s}=\frac{{{\mathbb{F}_q}\left[ v \right]}}{{\left\langle {{v^s} - v} \right\rangle }}$. For
further information about linear codes, cyclic codes and $m$-adic residue
codes the readers may consult the references \cite{A1,B1,MacWilliams}.

Let $q$ be a prime power and $\mathbb{F}_{q}$ be the finite field with $q$
elements. An $\left[ n,k\right] _{q}$ linear code $\mathcal{C}$ of length $n$
over $\mathbb{F}_{q}$ is a $k$-dimensional subspace of the vector space $%
\mathbb{F}_{q}^{n}.$ The elements of $\mathcal{C}$ are of the form $\left(
c_{0},c_{1},\ldots ,c_{n-1}\right) $ and are called codewords. The Hamming
weight of any $c\in \mathcal{C}$ is the number of nonzero coordinates of $c$
and is denoted by $w\left( c\right) .$ The minimum distance of $\mathcal{C}$ is
defined as $d=\min \left\{ \left. w\left( c\right) \right\vert \mathbf{0}%
\neq c\in \mathcal{C}\right\}$. The usual inner product $\left\langle {c,d} \right\rangle $
of two vectors $c=\left(c_{0},c_{1},\ldots ,c_{n-1}\right) $ and $d=\left(d_{0},d_{1},\ldots ,d_{n-1}\right)$
in $\mathbb{F}_{q}^{n}$ is defined to be $\left\langle {c,d} \right\rangle  = \sum\limits_{i = 0}^{n - 1} {{c_i}{d_i}} $ in $\mathbb{F}_{q}$.
The dual $C^\bot $ of the code $C$ of length $n$ over $\mathbb{F}_{q}$ is the set ${C^ \bot } = \left\{ {d \in \mathbb{F}_q^n:\left\langle {c,d} \right\rangle  = 0,\,\forall c \in C} \right\}$. Note that $C^\bot $ is an $\left[ n,n-k\right] _{q}$ linear code if $C$ is an $\left[ n,k\right] _{q}$ linear code.

A linear code $\mathcal{C}$ of length $n$
over $\mathbb{F}_{q}$ is said to be cyclic if for any codeword $%
(c_{0},c_{1},\ldots ,c_{n-1})\in \mathcal{C}$ we have that $%
(c_{n-1},c_{0},c_{1},\ldots ,c_{n-2})\in \mathcal{C}.$ A cyclic code of
length $n$ over $\mathbb{F}_{q}$ corresponds to a principal ideal $%
\left\langle g\left( x\right) \right\rangle $ of the quotient ring $\frac{{{\mathbb{F}_q}\left[ x \right]}}{{\left\langle {{x^n} - 1} \right\rangle }}$ where $\left.
g\left( x\right) \right\vert {{x^{n}-1}.}$ and $g\left( x\right)$ is called
the generator polynomial of the cyclic code $\mathcal{C}$ if $g\left(
x\right)$ is monic. If $\left( {n,q} \right) = 1$, there exits a primitive $%
n^{th}$ root of unity $\beta$ over some extension of $\mathbb{F}_q$. In this case, the
defining set for a cyclic code $\mathcal{C}=\left\langle g\left( x\right)
\right\rangle $ is the set $Z = \left\{ {i \in \left\{ {0,1, \ldots ,n - 1}
\right\}:g\left( {\beta ^i} \right) = 0} \right\}$. A polynomial $e\left( x \right) \in {\mathbb{F}_q}\left[ x \right]$ satisfying
${e^2}\left( x \right) = e\left( x \right)$ is called an idempotent polynomial. Recall that for any cyclic code $C$ over $\mathbb{F}_q$,
there exists unique idempotent generator $e\left( x \right)$ such that $C = \left\langle {e\left( x \right)} \right\rangle $. The dual $C^\bot $ is
a cyclic code generated by ${h^R}\left( x \right)$ of length $n$ over $\mathbb{F}_q$ if $C$ is a cyclic code generated by ${g}\left( x \right)$ of length $n$ over $\mathbb{F}_q$, where $g\left( x \right)h\left( x \right) = {x^n} - 1$ and the polynomial ${h^R}\left( x \right)$ is defined to be ${h^R}\left( x \right) = {x^{\deg h\left( x \right)}}h\left( {{x^{ - 1}}} \right)$ and is called the reciprocal polynomial of ${h}\left( x \right)$. Note that a cyclic code $C$ generated by ${g}\left( x \right)$ of length $n$ is of $n-k$ dimensional and in this case its dual $C^\bot $ is of $k$ dimensional.

In the following series of the definitions and examples we remind the
structure of the $m$-adic residue codes over finite fields.

\begin{definition}
\cite{A1} Let $p$ be a prime and $b$ be a primitive element of $\mathbb{Z}%
_{p}^{\ast }=\mathbb{Z}_{p}\setminus \{0\}$. The set of nonzero $m$-adic
residues modulo $p$ is defined as ${Q_0} = \left\{ {{a^m}:a \in \mathbb{Z}_p^ * } \right\}$
where $m\geqslant 2$, $m\in \mathbb{Z}$ and $m|(p-1).$ Moreover, set $Q_{i}=b^{i}Q_{0}$ for $0 \le i \le m - 1$.
\end{definition}

\begin{example}
Let $p=13$. Then $2$ is a primitive element of $\mathbb{Z}_{13}^{\ast }.$
Since $3|13-1$ we can take $m=3.$ Whence, we have $Q_{0}=\{1,5,8,12\},$ $%
Q_{1}=\{2,3,10,11\},$ $Q_{2}=\{4,6,7,9\}.$
\end{example}

\begin{definition}
\cite{A1} Let $p$ be a prime and $q$ be a prime power such that $p$ and $q$
are coprime. Let $b$ be a primitive element of $\mathbb{Z}_{p}^{\ast }$ and $%
\gamma $ be a primitive $p^{th}$ root of unity in some field extension of $%
\mathbb{F}_{q}.$ Let $Q_{0}$ be the set of nonzero $m$-adic residues modulo $%
p$ and $Q_{i}=b^{i}Q_{0}.$ If $q$ is an $m$-adic residue modulo $p,$ then
the codes generated by polynomials $g_{i}(x)=\frac{x^{p}-1}{\prod_{k\in
Q_{i}}{(x-\gamma ^{k})}}$ $(i=0,1,\ldots ,m-1)$ are called even-like
families of $m$-adic residue codes of class $I$ of length $p$ over $\mathbb{F%
}_{q}.$
\end{definition}

The other families of $m$-adic residue codes can be defined by derivations
of even-like family of $m$-adic residue codes of class $I.$

\begin{definition}
\cite{A1}

\begin{enumerate}
\item The codes generated by the polynomials $\widehat{g_{i}}(x)=\prod_{k\in
Q_{i}}{(x-\gamma ^{k})},$ where $(i=0,1,\ldots ,m-1),$ are called odd-like
class-$I$ $m$-adic residue codes of length $p$ over $\mathbb{F}_{q}$ and the
code generated by $\widehat{g_{i}}(x)$ is the complement of the code
generated by $g_{i}(x)$.

\item The codes generated by the polynomials ${h_{i}}(x)=(x-1)\widehat{g_{i}}%
(x),$ where $(i=0,1,\ldots ,m-1),$ are called even-like class $II$ $m$-adic
residue codes of length $p$ over $\mathbb{F}_{q}$ and the code generated by $%
h_{i}(x)$ is the complementary code of the code generated by $g_{i}(x)$.

\item The codes generated by the polynomials $\widehat{h_{i}}(x)=\frac{%
g_{i}(x)}{x-1},$ where $(i=0,1,\ldots ,m-1),$ are called odd-like class $II$
$m$-adic residue codes of length $p$ over $\mathbb{F}_{q}$ and these codes
are the complement of the codes generated by $h_{i}(x).$
\end{enumerate}
\end{definition}

In the following example, we list all even-like class-$I,$ $6$-adic residue
codes of length $19$ over $\mathbb{F}_7.$

\begin{example}
Even-like class $I,$ $6$-adic residue codes of length $19$ over $\mathbb{F}%
_{7}$ are
\begin{equation*}
\mathcal{C}_{0}=\langle
1+4x+6x^{2}+6x^{3}+3x^{4}+4x^{6}+5x^{7}+x^{8}+2x^{10}+2x^{11}+2x^{12}+4x^{13}+5x^{14}+3x^{15}+x^{16}\rangle ,
\end{equation*}%
\begin{equation*}
\mathcal{C}_{1}=\langle
1+3x+x^{2}+x^{3}+5x^{4}+x^{5}+6x^{6}+x^{7}+5x^{8}+6x^{9}+6x^{11}+3x^{12}+3x^{13}+5x^{14}+x^{15}+x^{16}\rangle ,
\end{equation*}%
\begin{equation*}
\mathcal{C}_{2}=\langle
1+5x^{2}+x^{3}+4x^{4}+3x^{5}+5x^{7}+3x^{8}+4x^{9}+6x^{10}+2x^{11}+6x^{12}+2x^{13}+4x^{14}+2x^{15}+x^{16}\rangle ,
\end{equation*}%
\begin{equation*}
\mathcal{C}_{3}=\langle
1+3x+5x^{2}+4x^{3}+2x^{4}+2x^{5}+2x^{6}+x^{8}+5x^{9}+4x^{10}+3x^{12}+6x^{13}+6x^{14}+4x^{15}+x^{16}\rangle ,
\end{equation*}%
\begin{equation*}
\mathcal{C}_{4}=\langle
1+x+5x^{2}+3x^{3}+3x^{4}+6x^{5}+6x^{7}+5x^{8}+x^{9}+6x^{10}+x^{11}+5x^{12}+x^{13}+x^{14}+3x^{15}+x^{16}\rangle ,
\end{equation*}%
and
\begin{equation*}
\mathcal{C}_{5}=\langle
1+2x+4x^{2}+2x^{3}+6x^{4}+2x^{5}+6x^{6}+4x^{7}+3x^{8}+5x^{9}+3x^{11}+4x^{12}+x^{13}+5x^{14}+x^{16}\rangle .
\end{equation*}%
The minimum distance of all these codes is $15.$ Hence, all these codes are
optimal.
\end{example}

Let $s>1$ be a natural number and $q$ be a prime such that $q\equiv
1(mod(s-1))$. Then we define the commutative ring $\mathcal{R}_{q,s}$ as
the quotient ring $\frac{{{\mathbb{F}_q}\left[ v \right]}}{{\left\langle {{v^s} - v} \right\rangle }}$. It is easy to see
that $\mathcal{R}_{q,s}\cong \mathbb{F}_{q}+v\mathbb{F}_{q}+v^{2}\mathbb{F}%
_{q}+\cdots +v^{s-1}\mathbb{F}_{q}.$

An element $\eta \in \mathcal{R}_{q,s}$ is called idempotent if $\eta
^{2}=\eta $ and two idempotents $\eta _{1},$ $\eta _{2}$ are said to be
orthogonal if $\eta _{1}\eta _{2}=0.$ An idempotent of $\mathcal{R}_{q,s}$ is
said to be primitive if it is non-zero and it can not be written as sum of
orthogonal idempotents. A collection $\left\{ \eta _{0},\eta _{1},\ldots ,\eta
_{s-1}\right\} $ of idempotents of $\mathcal{R}_{q,s}$ is complete if $\eta
_{0}+\eta_{1}+\ldots +\eta _{s-1}=1.$ Any complete collection of idempotents
in $\mathcal{R}_{q,s}$ is a basis of the $\mathbb{F}_{q}$-vector space $%
\mathcal{R}_{q,s}.$ Hence, any element $r\in \mathcal{R}_{q,s}$ can be
uniquely represented as $r=r_{0}\eta _{0}+r_{1}\eta _{1}+\ldots +r_{s-1}\eta
_{s-1},$ where $r_{i}\in \mathcal{R}_{q,s}.$

If $\alpha $ is a primitive element of $\mathbb{F}_{q}^{\ast },$ then $\zeta
=\alpha ^{\frac{q-1}{s-1}}$ is a primitive $(s-1)^{th}$ root of unity and $%
1+\zeta +\zeta ^{2}+\cdots +\zeta ^{s-2}=0.$ Let $\eta _{0},\eta _{1},\ldots
,\eta _{s-1}$ denote the following elements of $\mathcal{R}_{q,s}.$

\begin{eqnarray}
\eta _{0} &=&1-v^{s-1}  \nonumber \\
\eta _{1} &=&(s-1)^{-1}(v+v^{2}+\cdots +v^{s-2}+v^{s-1})  \nonumber \\
\eta _{2} &=&(s-1)^{-1}(\zeta v+\zeta ^{2}v^{2}+\cdots +\zeta
^{s-2}v^{s-2}+v^{s-1})  \nonumber \\
\eta _{3} &=&(s-1)^{-1}((\zeta ^{2})v+(\zeta ^{2})^{2}v^{2}+\cdots +(\zeta
^{2})^{s-2}v^{s-2}+v^{s-1})  \nonumber \\
\vdots   \nonumber \\
\eta _{s-1} &=&(s-1)^{-1}((\zeta ^{s-2})v+(\zeta ^{s-2})^{2}v^{2}+\cdots
+(\zeta ^{s-2})^{s-2}v^{s-2}+v^{s-1}).  \nonumber
\end{eqnarray}

It can be easily seen that $\eta _{i}^{2}=\eta _{i},$ $\eta _{i}\eta _{j}=0$
for all $i$ and $j$ such that $0\leq i,j\leq s-1$, $i\neq j$. Further, $%
\sum\limits_{i=0}^{s-1}{\eta _{i}=1}$. By the decomposition theorem of ring
theory, we can decompose $\mathcal{R}_{q,s}$ as $\mathcal{R}_{q,s}=\eta _{0}%
\mathcal{R}_{q,s}\oplus \eta _{1}\mathcal{R}_{q,s}\oplus \cdots \oplus \eta
_{s-1}\mathcal{R}_{q,s}.$ See that $\left\langle {{\eta _{i}}}\right\rangle
=\left\{ {\lambda {\eta _{i}}:\lambda \in {\mathbb{F}_{q}}}\right\} $.
Hence, $\mathcal{R}_{q,s}={\eta _{0}}{\mathbb{F}_{q}}\oplus {\eta _{1}}{%
\mathbb{F}_{q}}\oplus \cdots \oplus {\eta _{s-1}}{\mathbb{F}_{q}}$, which
implies that every element $r$ of the ring $\mathcal{R}_{q,s}$ is written
uniquely as the sum $r={\lambda _{0}}{\eta _{0}}+{\lambda _{1}}{\eta _{1}}%
+\cdots +{\lambda _{s-1}}{\eta _{s-1}}$ for some ${\lambda _{0}},{\lambda
_{1}},\ldots ,{\lambda _{s-1}}\in {\mathbb{F}_{q}}$.

Let $\mathcal{R}_{q,s}^n$ be the set of $n$-tuples over $\mathcal{R}_{q,s}$. Then, $\mathcal{R}_{q,s}^n$ is an $\mathcal{R}_{q,s}$-module. A nonempty subset of $\mathcal{R}_{q,s}^n$ is a code over $\mathcal{R}_{q,s}$ of length $n$. An $\mathcal{R}_{q,s}$-submodule of $\mathcal{R}_{q,s}^n$ is a linear code over $\mathcal{R}_{q,s}$ of length $n$. For a code $C$ over $\mathcal{R}_{q,s}$ of length $n$, its dual $C^{\bot}$ is the set ${C^ \bot } = \left\{ {y \in \mathcal{R}_{q,s}^n:\left\langle {x,y} \right\rangle  = 0,\,\forall x \in C} \right\}$. Note that if $C$ is a linear code over $\mathcal{R}_{q,s}$ of length $n$, then its dual is also linear code over $\mathcal{R}_{q,s}$ of length $n$.

\section{$m$-adic residue codes over $\mathcal{R}_{q,s}$}

In \cite{A2}, Pless introduced $m$-adic residue codes in terms of idempotent
generators and gave relation between these idempotent generators. Kuruz et
al. studied $m$-adic residue codes over the ring $\mathcal{R}_{q,2} = \frac{{{\mathbb{F}_q}\left[ v \right]}}{{\left\langle {{v^2} - v} \right\rangle }}$ (see
\cite{kuruz}) and identified the idempotent generators of these codes. Also,
they gave some conditions for the generator polynomials to be
palindromic.

In this section we generalize idempotent generators of the $m$-adic residue
codes to the ring $\mathcal{R}_{q,s}$ and we identify the idempotent
generators of all classes of the $m$-adic residue codes. Furthermore, we
determine the algebraic structure for each class of $m$-adic residue codes
over $\mathcal{R}_{q,s}.$

\begin{proposition}\label{prop01}
Let $p$ be a prime and $q$ be a prime power. Let $m$ be a positive integer
dividing $p-1$. Let $q$ be an $m$-adic residue modulo $p,$
i.e., $q\in Q_{0}$. If, for each $0\leqslant i\leqslant s-1$, $f_{i}$ is idempotent in $\frac{{{\mathbb{F}_q}\left[ x \right]}}{{\left\langle {{x^p} - 1} \right\rangle }}$, then $\sum\limits_{i=0}^{s-1}{\eta _{i}f_{i}}$ is idempotent in $\frac{{{\mathcal{R}
_{q,s}}\left[ x \right]}}{{\left\langle {{x^p} - 1} \right\rangle }}$.
\end{proposition}

\begin{proof}
Since $\eta _{i}^{2}=\eta _{i}$ and $\eta _{i}\eta _{j}=0$ for all $0\leq i,j\leq s-1$ where $i\neq j$, we have $(\sum\limits_{i=0}^{s-1}{%
\eta _{i}f_{i}})^{2}=\sum\limits_{i=0}^{s-1}{(\eta _{i}f_{i})^{2}}%
=\sum\limits_{i=0}^{s-1}{\eta _{i}f_{i}}$.
\end{proof}

It follows from Proposition \ref{prop01} that $E=\sum\limits_{i=0}^{s-1}{\eta _{i}e_{i}}$ is idempotent when $e_{i}$ is idempotent generator of an even-like
class-$I$ $m$-adic residue code over $\mathbb{F}_q$ and so we give the following definition of even-like class-$I$ $m$-adic residue
codes of length $p$ over $\mathcal{R}_{q,s}$ with respect to their idempotent generators.

\begin{definition}
For each $0 \le i \le s - 1$, let $e_i$ be the idempotent generator of an even-like class-$I$
$m$-adic residue code of length $p$ over $\mathbb{F}_q$. A code over $\mathcal{R}_{q,s}$ generated by
$E=\sum\limits_{i=0}^{s-1}{\eta _{i}e_{i}}$ is called an even-like class-$I$ $m$-adic residue code
of length $p$ over $\mathcal{R}_{q,s}$.
\end{definition}

We give the following characterization for even-like class-$I$ $m$-adic
residue codes of length $p$ over $\mathcal{R}_{q,s}.$

\begin{theorem}
\label{thr1} For any even-like class-$I$ $m$-adic residue code $\mathcal{C}$
of length $p$ over $\mathcal{R}_{q,s},$ there exists even-like class-$I$ $m$%
-adic residue codes $\mathcal{C}{_{0}},\mathcal{C}{_{1}},\ldots ,\mathcal{C}{%
_{s-1}}$ of length $p$ over $\mathbb{F}_{q}$ with generator polynomials ${%
g_{0}}\left( x\right) ,{g_{1}}\left( x\right) ,\ldots ,{g_{s-1}}\left(
x\right) $ such that $\mathcal{C}=\left\langle {g\left( x\right) }%
\right\rangle ={\eta _{0}}\mathcal{C}_{0}\oplus {\eta }_{1}\mathcal{C}%
_{1}\oplus \cdots \oplus {\eta _{s-1}}\mathcal{C}_{s-1}$ where $g\left(
x\right) ={\eta _{0}}{g_{0}}\left( x\right) +{\eta _{1}}{g_{1}}\left(
x\right) +\cdots +{\eta _{s-1}}{g_{s-1}}\left( x\right) $ and $\left\vert
\mathcal{C}\right\vert ={q^{ps-\sum\limits_{i=0}^{s-1}{\deg {g_{i}}\left(
x\right) }}}.$
\end{theorem}

\begin{proof}
By the definition, we take an even-like class-$I$ $m$-adic residue code $%
\mathcal{C}$ over $\mathcal{R}_{q,s}$ as a code generated by $%
E=\sum\limits_{i=0}^{s-1}{\eta _{i}e_{i}}$. Define the set $J=\left\langle {{%
\eta _{0}}{e_{0}},{\eta _{1}}{e_{1}},\ldots ,{\eta _{s-1}}{e_{s-1}}}%
\right\rangle .$ Then, clearly $\mathcal{C}\subseteq J$. Conversely, for all
$0\leq i\neq j\leq s-1$, it follows from ${\eta _{i}}{\eta _{j}}=0$ and ${%
\eta _{i}}{\eta _{i}}={\eta _{i}}$ that ${\eta _{i}}\left( {{\eta _{0}}{e_{0}%
}+\cdots +{\eta _{i}}{e_{i}}+\cdots +{\eta _{s-1}}{e_{s-1}}}\right) ={\eta
_{i}}{e_{i}}$ and so $J\subseteq \mathcal{C}.$ Hence, $J=\mathcal{C}.$ The
last equality implies that $\mathcal{C}={\eta _{0}}\mathcal{C}_{0}\oplus {%
\eta }_{1}\mathcal{C}_{1}\oplus \cdots \oplus {\eta _{s-1}}\mathcal{C}_{s-1}$
where $\mathcal{C}_{i}$ is an even-like class-$I$ $m$-adic residue code over
$\mathbb{F}_{q}$ with idempotent generator $e_{i}.$ Letting ${g_{i}}\left(
x\right) $ to be generator polynomial for $\mathcal{C}_{i}$, one gets that $%
g\left( x\right) ={\eta _{0}}{g_{0}}\left( x\right) +{\eta _{1}}{g_{1}}%
\left( x\right) +\cdots +{\eta _{s-1}}{g_{s-1}}\left( x\right) $ and $%
\left\vert \mathcal{C}\right\vert ={q^{ps-\sum\limits_{i=0}^{s-1}{\deg {g_{i}%
}\left( x\right) }}}.$
\end{proof}

We have from \cite{A4} that $1-e$ is idempotent generator of an odd-like class-$I$ $m$-adic residue code over $\mathbb{F}_q$ when $e$ is idempotent generator of an even-like class-$I$ $m$-adic residue code over $\mathbb{F}_q$. Then, Proposition \ref{prop01} implies that $E^{\prime}=\sum\limits_{i=0}^{s-1}{\eta _{i}\left (1-e_{i} \right )}=1-E$ is idempotent if $e_{i}$ is idempotent generator of an even-like
class-$I$ $m$-adic residue code over $\mathbb{F}_q$. Hence, we consider the odd-like class-$I$ $m$-adic residue
codes of length $p$ over $\mathcal{R}_{q,s}$ with respect to their idempotent generators.

\begin{definition}
Let $E=\sum\limits_{i=0}^{s-1}{\eta _{i}e_{i}}$ where for $0 \le i \le s - 1$, $e_i$ is
the idempotent generator of an even-like class-$I$ $m$-adic residue code of length $p$ over $\mathbb{F}_q$.
A code over $\mathcal{R}_{q,s}$ generated by $E^{\prime}=1-E=\sum\limits_{i=0}^{s-1}{\eta _{i}\left (1-e_{i} \right )}$ is called an odd-like class-$I$ $m$-adic residue code
of length $p$ over $\mathcal{R}_{q,s}$.
\end{definition}

Similar to the characterization for even-like class-$I$ $m$-adic residue
codes, we characterize odd-like class-$I$ $m$-adic residue codes of length $p
$ over $\mathcal{R}_{q,s}$ in the following theorem.

\begin{theorem}
\label{thr2} For any odd-like class-$I$ $m$-adic residue code $\mathcal{C}$
of length $p$ over $\mathcal{R}_{q,s},$ there exists odd-like class-$I$ $m$%
-adic residue codes $\mathcal{C}{_{0}},\mathcal{C}{_{1}},\ldots ,\mathcal{C}{%
_{s-1}}$ of length $p$ over $\mathbb{F}_{q}$ with generator polynomials ${%
g_{0}}\left( x\right) ,{g_{1}}\left( x\right) ,\ldots ,{g_{s-1}}\left(
x\right) $ such that $\mathcal{C}=\left\langle {g\left( x\right) }%
\right\rangle ={\eta _{0}}\mathcal{C}_{0}\oplus {\eta }_{1}\mathcal{C}%
_{1}\oplus \cdots \oplus {\eta _{s-1}}\mathcal{C}_{s-1}$ where $g\left(
x\right) ={\eta _{0}}{g_{0}}\left( x\right) +{\eta _{1}}{g_{1}}\left(
x\right) +\cdots +{\eta _{s-1}}{g_{s-1}}\left( x\right) $ and $\left\vert
\mathcal{C}\right\vert ={q^{ps-\sum\limits_{i=0}^{s-1}{\deg {g_{i}}\left(
x\right) }}}.$
\end{theorem}

\begin{proof}
Let $\mathcal{C}$ be an odd-like class-$I$ $m$-adic residue code $\mathcal{C}
$ of length $p$ over $\mathcal{R}_{q,s}.$ By the definition for odd-like
class-$I$ $m$-adic residue codes over $\mathcal{R}_{q,s},$ $\mathcal{C}%
=\left\langle {1-{E}}\right\rangle $. See that $1-E
=\sum\limits_{i=0}^{s-1}{{\eta _{i}}-\sum\limits_{i=0}^{s-1}{{\eta _{i}}{%
e_{i}}}=\sum\limits_{i=0}^{s-1}{{\eta _{i}}\left( {1-{e_{i}}}\right) }}$ and
define the set
\begin{equation*}
J=\left\langle {{\eta _{0}}\left( {1-{e_{0}}}\right) ,{\eta _{1}}\left( {1-{%
e_{1}}}\right) ,\ldots ,{\eta _{s-1}}\left( {1-{e_{s-1}}}\right) }%
\right\rangle .
\end{equation*}%
Then, it is clear that $\mathcal{C}\subseteq J$. Moreover, for each $%
0\leqslant i\leqslant s-1$, ${\eta _{i}}\sum\limits_{i=0}^{s-1}{{\eta _{i}}%
\left( {1-{e_{i}}}\right) }=\eta _{i}^{2}\left( {1-{e_{i}}}\right) ={\eta
_{i}}\left( {1-{e_{i}}}\right)\in \mathcal{C}$, we get $J\subseteq \mathcal{C}$ and so $J=\mathcal{C}.
$ Note that $1-e_{i}$ is an idempotent generator for an odd-like class-$I$ $m
$-adic residue code over $\mathbb{F}_{q}$ whenever $e_{i}$ is an idempotent
generator for an even-like class-$I$ $m$-adic residue code over $\mathbb{F}%
_{q}.$ Hence, the equality $J=\mathcal{C}$ implies that $\mathcal{C}={\eta
_{0}}\mathcal{C}_{0}\oplus {\eta }_{1}\mathcal{C}_
{1}\oplus \cdots \oplus {%
\eta _{s-1}}\mathcal{C}_{s-1}$ where $\mathcal{C}$ is an odd-like class-$I$ $%
m$-adic residue code over $\mathbb{F}_{q}$ with idempotent generator $%
1-e_{i}.$ Taking ${g_{i}}\left( x\right) $ as a generator polynomial of $%
\mathcal{C}_{i},$ one obtains that $g\left( x\right) ={\eta _{0}}{g_{0}}%
\left( x\right) +{\eta _{1}}{g_{1}}\left( x\right) +\cdots +{\eta _{s-1}}{%
g_{s-1}}\left( x\right) $ and $\left\vert \mathcal{C}\right\vert ={%
q^{ps-\sum\limits_{i=0}^{s-1}{\deg {g_{i}}\left( x\right) }}}.$
\end{proof}

Let $h(x)=x+x^{2}+\ldots +x^{p-1}.$ Note that $1-\frac{1}{p}h-e$ is
idempotent generator of an even-like class-$II$ $m$-adic residue code of
length $p$ over $\mathbb{F}_{q}$ when $e$ is idempotent generator of an
even-like class-$II$ $m$-adic residue code of length $p$ over $\mathbb{F}_{q}
$ \cite{A4}. In this case, from Proposition \ref{prop01}, we get that $%
D=\sum\limits_{i=0}^{s-1}{\eta _{i}\left( 1-\frac{1}{p}h-e_{i}\right) }=1-%
\frac{1}{p}h-E$ is idempotent if $e_{i}$ is idempotent generator of an
even-like class-$I$ $m$-adic residue code over $\mathbb{F}_{q}.$ Hence, we
consider the even-like class-$II$ $m$-adic residue codes of length $p$ over $%
\mathcal{R}_{q,s}$ with respect to their idempotent generators.

\begin{definition}
Let $E=\sum\limits_{i=0}^{s-1}{\eta _{i}e_{i}}$ where for $0 \le i \le s - 1$, $e_i$ is
the idempotent generator of an even-like class-$I$ $m$-adic residue code of length $p$ over $\mathbb{F}_q$.
A code over $\mathcal{R}_{q,s}$ generated by $D=1-\frac{1}{p}h-E=\sum\limits_{i=0}^{s-1}{\eta _{i}\left(1-\frac{1}{p}h-e_{i}\right)}$ is called an even-like class-$II$ $m$-adic residue code
of length $p$ over $\mathcal{R}_{q,s}$.
\end{definition}

The following is the exact structure of even-like class-$II$ $m$-adic
residue codes of length $p$ over $\mathcal{R}_{q,s}.$

\begin{theorem}
\label{thr3} Let $\mathcal{C}$ be an even-like class-$II$ $m$-adic residue
code of length $p$ over $\mathcal{R}_{q,s}.$ Then, for some even-like class-$%
II$ $m$-adic residue codes $\mathcal{C}{_{0}},\mathcal{C}{_{1}},\ldots ,%
\mathcal{C}{_{s-1}}$ of length $p$ over $\mathbb{F}_{q}$ with generator
polynomials ${g_{0}}\left( x\right) ,{g_{1}}\left( x\right) ,\ldots ,{g_{s-1}%
}\left( x\right) ,$ $\mathcal{C}=\left\langle {g\left( x\right) }%
\right\rangle ={\eta _{0}}\mathcal{C}_{0}\oplus {\eta }_{1}\mathcal{C}%
_{1}\oplus \cdots \oplus {\eta _{s-1}}\mathcal{C}_{s-1}$ where $g\left(
x\right) ={\eta _{0}}{g_{0}}\left( x\right) +{\eta _{1}}{g_{1}}\left(
x\right) +\cdots +{\eta _{s-1}}{g_{s-1}}\left( x\right) $ and $\left\vert
\mathcal{C}\right\vert ={q^{ps-\sum\limits_{i=0}^{s-1}{\deg {g_{i}}\left(
x\right) }}}$.
\end{theorem}

\begin{proof}
Let $\mathcal{C}$ be an even-like class-$II$ $m$-adic residue code of length
$p$ over $\mathcal{R}_{q,s}.$ By the definition for an even-like class-$II$ $%
m$-adic residue codes over $\mathcal{R}_{q,s},$ $\mathcal{C}=\left\langle {%
1-\frac{1}{p}h-{E}}\right\rangle $. See that $1-\frac{1}{p}h-E=\sum\limits_{i=0}^{s-1}{{{%
\eta _{i}}\left( 1-\frac{1}{p}h\right) }-\sum\limits_{i=0}^{s-1}{{\eta _{i}}{e_{i}}}%
=\sum\limits_{i=0}^{s-1}{{\eta _{i}}\left( {1-\frac{1}{p}h-{e_{i}}}\right) }}$ and
define the set $J=\left\langle {{\eta _{0}}\left( {1-\frac{1}{p}h-{e_{0}}}\right) ,{%
\eta _{1}}\left( {1-\frac{1}{p}h-{e_{1}}}\right) ,\ldots ,{\eta _{s-1}}\left( {1-\frac{1}{p}h-{%
e_{s-1}}}\right) }\right\rangle $. Then, it is clear that $\mathcal{C}%
\subseteq J$. Also, since ${\eta _{i}}\sum\limits_{i=0}^{s-1}{{\eta _{i}}%
\left( {1-\frac{1}{p}h-{e_{i}}}\right) }=\eta _{i}^{2}\left( {1-\frac{1}{p}h-{e_{i}}}\right) ={%
\eta _{i}}\left( {1-\frac{1}{p}h-{e_{i}}}\right)\in \mathcal{C}$ for each $0\leqslant i\leqslant s-1,$ we get $%
J\subseteq \mathcal{C},$ and so $J=\mathcal{C}.$ Note that $1-\frac{1}{p}h-e_{i}$ is
idempotent generator for an even-like class-$II$ $m$-adic residue code over $%
\mathbb{F}_{q}$ whenever $e_{i}$ is idempotent generator for an even-like
class-$I$ $m$-adic residue code over $\mathbb{F}_{q}$. Hence, it follows
from $J=\mathcal{C}$ that $\mathcal{C}={\eta _{0}}\mathcal{C}_{0}\oplus {%
\eta }_{1}\mathcal{C}_{1}\oplus \cdots \oplus {\eta _{s-1}}\mathcal{C}_{s-1}$
where $\mathcal{C}_{i}$ is an even-like class-$II$ $m$-adic residue code
over $\mathbb{F}_{q}$ with idempotent generator $1-\frac{1}{p}h-e_{i}.$ Let ${g_{i}}%
\left( x\right) $ to be generator polynomial for $\mathcal{C}_{i}.$ Then, it
is easy to see that $g\left( x\right) ={\eta _{0}}{g_{0}}\left( x\right) +{%
\eta _{1}}{g_{1}}\left( x\right) +\cdots +{\eta _{s-1}}{g_{s-1}}\left(
x\right) $ and $\left\vert \mathcal{C}\right\vert ={q^{ps-\sum%
\limits_{i=0}^{s-1}{\deg {g_{i}}\left( x\right) }}}.$
\end{proof}

We have from \cite{A4} that $\frac{1}{p}h+e$ is idempotent generator of an odd-like class-$II$ $m$-adic residue code over $\mathbb{F}_q$ when $e$ is idempotent generator of an even-like class-$I$ $m$-adic residue code over $\mathbb{F}_q$. Then, from Proposition \ref{prop01}, we have that $D^{\prime}=\sum\limits_{i=0}^{s-1}{\eta _{i}\left (\frac{1}{p}h+e_{i} \right )}=1-D=\frac{1}{p}h+E$ is idempotent if $e_{i}$ is idempotent generator of an even-like class-$I$ $m$-adic residue code over $\mathbb{F}_q$. Hence, we consider the odd-like class-$II$ $m$-adic residue
codes of length $p$ over $\mathcal{R}_{q,s}$ with respect to their idempotent generators.

\begin{definition}
Let $E=\sum\limits_{i=0}^{s-1}{\eta _{i}e_{i}}$ where for $0 \le i \le s - 1$, $e_i$ is
the idempotent generator of an even-like class-$I$ $m$-adic residue code of length $p$ over $\mathbb{F}_q$.
A code over $\mathcal{R}_{q,s}$ generated by $D^{\prime}=1-D=\frac{1}{p}h+E=\sum\limits_{i=0}^{s-1}{\eta _{i}\left (\frac{1}{p}h+e_{i} \right )}$ is called an odd-like class-$II$ $m$-adic residue code of length $p$ over $\mathcal{R}_{q,s}$.
\end{definition}

In the following, we give the exact structure for odd-like class-$II$ $m$%
-adic residue codes of length $p$ over $\mathcal{R}_{q,s}.$

\begin{theorem}
\label{thr33} Let $\mathcal{C}$ be an odd-like class-$II$ $m$-adic residue
code of length $p$ over $\mathcal{R}_{q,s}.$ Then, for some odd-like class-$%
II$ $m$-adic residue codes $\mathcal{C}{_{0}},\mathcal{C}{_{1}},\ldots ,%
\mathcal{C}{_{s-1}}$ of length $p$ over $\mathbb{F}_{q}$ with generator
polynomials ${g_{0}}\left( x\right) ,{g_{1}}\left( x\right) ,\ldots ,{g_{s-1}%
}\left( x\right) ,$ $\mathcal{C}=\left\langle {g\left( x\right) }%
\right\rangle ={\eta _{0}}\mathcal{C}_{0}\oplus {\eta }_{1}\mathcal{C}%
_{1}\oplus \cdots \oplus {\eta _{s-1}}\mathcal{C}_{s-1}$ where $g\left(
x\right) ={\eta _{0}}{g_{0}}\left( x\right) +{\eta _{1}}{g_{1}}\left(
x\right) +\cdots +{\eta _{s-1}}{g_{s-1}}\left( x\right) $ and $\left\vert
\mathcal{C}\right\vert ={q^{ps-\sum\limits_{i=0}^{s-1}{\deg {g_{i}}\left(
x\right) }}}.$
\end{theorem}

\begin{proof}
Let $\mathcal{C}$ be an odd-like class-$II$ $m$-adic residue code of length $%
p$ over $\mathcal{R}_{q,s}.$ By the definition for an odd-like class-$II$ $m$%
-adic residue codes over $\mathcal{R}_{q,s},$ $\mathcal{C}=\left\langle {\frac{1}{p}h+{
E}}\right\rangle$. See that $\frac{1}{p}h+{E}=\sum\limits_{i=0}^{s-1}{{{\eta
_{i}}\frac{1}{p}h}+\sum\limits_{i=0}^{s-1}{{\eta _{i}}{e_{i}}}=\sum\limits_{i=0}^{s-1}{{%
\eta _{i}}\left( {\frac{1}{p}h+{e_{i}}}\right) }}$ and define the set
\begin{equation*}
J=\left\langle {{\eta _{0}}\left( {\frac{1}{p}h+{e_{0}}}\right) ,{\eta _{1}}\left( {\frac{1}{p}h+{%
e_{1}}}\right) ,\ldots ,{\eta _{s-1}}\left( {\frac{1}{p}h+{e_{s-1}}}\right) }%
\right\rangle .
\end{equation*}%
Then, it is clear that $\mathcal{C}\subseteq J.$ Also, since ${\eta _{i}}%
\sum\limits_{i=0}^{s-1}{{\eta _{i}}\left( {\frac{1}{p}h+{e_{i}}}\right) }=\eta
_{i}^{2}\left( {\frac{1}{p}h+{e_{i}}}\right) ={\eta _{i}}\left( {\frac{1}{p}h+{e_{i}}}\right)\in \mathcal{C}$ for each $%
0\leq i\leq s-1,$ we get $J\subseteq \mathcal{C},$ and so $J=\mathcal{C}.$
Note that $\frac{1}{p}h+e_{i}$ is idempotent generator for an odd-like class-$II$ $m$%
-adic residue code over $\mathbb{F}_{q}$ whenever $e_{i}$ is idempotent
generator for an even-like class-$I$ $m$-adic residue code over $\mathbb{F}%
_{q}$. Hence, it follows from $J=\mathcal{C}$ that $\mathcal{C}={\eta _{0}}%
\mathcal{C}_{0}\oplus {\eta }_{1}\mathcal{C}_{1}\oplus \cdots \oplus {\eta
_{s-1}}\mathcal{C}_{s-1}$ where $\mathcal{C}_{i}$ is an odd-like class-$II$ $%
m$-adic residue code over $\mathbb{F}_{q}$ with idempotent generator $\frac{1}{p}h+e_{i}
$. Let ${g_{i}}\left( x\right) $ be the generator polynomial of $\mathcal{C}%
_{i}.$ Then, it is easy to see that $g\left( x\right) ={\eta _{0}}{g_{0}}%
\left( x\right) +{\eta _{1}}{g_{1}}\left( x\right) +\cdots +{\eta _{s-1}}{%
g_{s-1}}\left( x\right) $ and $\left\vert \mathcal{C}\right\vert ={%
q^{ps-\sum\limits_{i=0}^{s-1}{\deg {g_{i}}\left( x\right) }}}.$
\end{proof}

In the following example, we list some of even-like
class-$I$ $4$-adic residue codes of length $13$ over $\mathcal{R}_{3,3}$.

\begin{example}
The idempotent generators of even-like class-$I$ $4$-adic residue codes of
length $13$ over $\mathbb{F}_{3}$ are $e_{0}=l_{0}+2l_{2}+2l_{3},$ $%
e_{1}=2l_{1}+2l_{2}+l_{3},$ $e_{2}=2l_{0}+2l_{1}+l_{2}$ and $%
e_{3}=2l_{0}+l_{1}+2l_{3},$ where $l_{0}=x+x^{3}+x^{9},$ $%
l_{1}=x^{2}+x^{5}+x^{6},$ $l_{2}=x^{4}+x^{10}+x^{12},$ $%
l_{3}=x^{7}+x^{8}+x^{11}.$ Let $E_{0}=\eta _{0}e_{0}+\eta_{1}e_{1}+\eta _{2}e_{2}$,
$E_{1}=\eta _{0}e_{3}+\eta _{1}e_{0}+\eta _{2}e_{1}$,
$E_{2}=\eta _{0}e_{2}+\eta _{1}e_{3}+\eta _{2}e_{0}$,
$E_{3}=\eta _{0}e_{1}+\eta _{1}e_{2}+\eta _{2}e_{3}$, where $\eta _{0}=1-v^{2}$, $\eta _{1}=2v+2v^{2}$ and $\eta _{2}=v+2v^{2}$.
\newline
Let $g_{i}(x)$ be the generator polynomial of even-like class-$I$ $4$-adic residue code of length $13$ over $\mathcal{R}_{3,3}$ with idempotent
generator $E_{i}$, then \newline
$%
g_{0}(x)=1+(2v+1)x+(2v+2v^{2})x^{2}+(2v^{2}+v+1)x^{3}+(v^{2}+2v+1)x^{4}+(2v^{2}+2v+1)x^{5}+2vx^{6}+(v+2)x^{7}+vx^{8}+(v^{2}+2v+2)x^{9}+x^{10},
$\newline
$%
g_{1}(x)=1+2vx+(2v+1)x^{2}+(v+1)x^{3}+(2v+1)x^{4}+(2v+2)x^{5}+(2v+2)x^{6}+vx^{7}+(v+1)x^{8}+(2v+2)x^{9}+x^{10},
$\newline
$g_{2}(x)=g_{3}(x)=1+x+2x^{2}+2x^{3}+2x^{5}+2x^{8}+x^{9}.$\newline
The parameters of the codes generated by $g_{0}(x),$ $g_{1}(x),$ $g_{2}(x)$
and $g_{3}(x)$ are $[13,3,9],$ $[13,3,6],$ $[13,4,6],$ and $[13,4,6]$
respectively. The code generated by $g_{0}(x)$ attains the Griesmer bound
given in \cite{shiro}. The other codes are nearly optimal with respect to
Griesmer bound for rings.
\end{example}

\section{Quantum Codes from m-adic Residue Codes over $\mathcal{R}_{q,s}$}

In this section, as an application of m-adic residue codes over the ring $%
R_{q,s},$ we are going to derive some families of quantum error correcting
codes. Before giving the well-known quantum code construction, called CSS
construction, from linear codes over finite fields, we here give a brief
introduction for quantum codes.

An ${\left[ \kern-0.15em\left[ {n,k,d}\right] \kern-0.15em\right] _{q}}$
quantum code is a $q^{k}$-dimensional subspace of $q^{n}$-dimensional
complex vector space ${\mathbb{C}^{{\ \otimes ^{{q^{n}}}}}},$ where $\otimes
$ denotes the tensor product and, $n,$ $k$ and $d$ are said to be the
parameters of the quantum code. Such a quantum code encodes $k$ qudits to $n$
qudits and corrects up to $\left\lfloor {\frac{{d-1}}{2}}\right\rfloor $
errors. There exists an analogue between the linear codes over finite fields
and quantum codes as follows:

\begin{theorem}[CSS Code Construction]
\label{thr0}\cite{ketkar} Let $\mathcal{C}_{1}$ and $\mathcal{C}_{2}$ be two
linear codes over the finite field $\mathbb{F}_{q}$ having the parameters ${%
\left[ {n,{k_{1}},{d_{1}}}\right] _{q}}$ and ${\left[ {n,{k_{2}},{d_{2}}}%
\right] _{q}}$ such that $\mathcal{C}_{2}^{\bot }\subseteq \mathcal{C}{_{1}}$%
. Then, there exists an ${\left[ \kern-0.15em\left[ {n,{k_{1}}+{k_{2}}-n,d}%
\right] \kern-0.15em\right] _{q}}$ quantum error correcting code where
\begin{equation*}
d=\min \left\{ {w\left( c\right) :c\in \left( \mathcal{C}{{_{1}}-\mathcal{C}%
_{2}^{\bot }}\right) \cup \left( \mathcal{C}{{_{2}}-\mathcal{C}_{1}^{\bot }}%
\right) }\right\} .
\end{equation*}
In particular, if $\mathcal{C}_{1}^{\bot }\subseteq \mathcal{C}{_{1}},$ then
there exits an ${\left[ \kern-0.15em\left[ {n,2k_{1}-n,{d_{1}}}\right] \kern%
-0.15em\right] _{q}}$ quantum error correcting code.
\end{theorem}

We determine the structure of the duals for each class of $m$-adic residue
codes over $\mathcal{R}_{q,s}.$

\begin{theorem}
\label{thr4} Let $\mathcal{C}{_{0}},\mathcal{C}{_{1}},\ldots ,\mathcal{C}{%
_{s-1}}$ be $m$-adic residue codes of length $p$ over $\mathbb{F}_{q}$ with
the generator polynomials ${g_{0}}\left( x\right) ,{g_{1}}\left( x\right)
,\ldots ,{g_{s-1}}\left( x\right) $, respectively and let $\mathcal{C}={\eta
_{0}}\mathcal{C}_{0}\oplus {\eta }_{1}\mathcal{C}_{1}\oplus \cdots \oplus {%
\eta _{s-1}}\mathcal{C}_{s-1}$ be an $m$-adic residue code of length $p$
over $\mathcal{R}_{q,s}.$ Then, the dual code of $\mathcal{C}$ is $\mathcal{C%
}{^{\bot }}=\left\langle {{g^{\bot }}\left( x\right) }\right\rangle ={\eta
_{0}}\mathcal{C}_{0}^{\bot }\oplus {\eta _{1}}\mathcal{C}_{1}^{\bot }\oplus
\cdots \oplus {\eta _{s-1}}\mathcal{C}_{s-1}^{\bot }$ where $\mathcal{C}%
_{i}^{\bot }$'s are the duals of $\mathcal{C}_{i}$'s over $\mathbb{F}_{q},$ $%
{g^{\bot }}\left( x\right) ={\eta _{0}}h_{0}^{R}\left( x\right) +{\eta _{1}}%
h_{1}^{R}\left( x\right) +\cdots +{\eta _{s-1}}h_{s-1}^{R}$ and ${x^{p}}-1={%
g_{i}}\left( x\right) {h_{i}}\left( x\right) $ over $\mathbb{F}_{q}$ for
each $i$.
\end{theorem}

\begin{proof}
Define the ideal $J=\left\langle {{\eta _{0}}h_{0}^{R}\left( x\right) ,{\eta
_{1}}h_{1}^{R}\left( x\right) ,\ldots ,{\eta _{s-1}}h_{s-1}^{R}}%
\right\rangle $. Since ${\eta _{i}}{g_{i}}\left( x\right) {\eta _{j}}{h_{j}}%
\left( x\right) =0$ for all $0\leqslant i,j\leqslant s-1,$ $J\subseteq
\mathcal{C}^{\bot }.$ It follows from comparing the cardinalities of $J$ and
$\mathcal{C}^{\bot }$ that $J=\mathcal{C}^{\bot }.$ Then, it is not
difficult to see that $\mathcal{C}^{\bot }$ is a code generated by the
polynomial ${g^{\bot }}\left( x\right) ={\eta _{0}}h_{0}^{R}\left( x\right) +%
{\eta _{1}}h_{1}^{R}\left( x\right) +\cdots +{\eta _{s-1}}h_{s-1}^{R}$.
\end{proof}

Now, we derive a condition for $m$-adic residue codes over $R_{q,s}$ to
contain their duals.

\begin{proposition}
\label{prop1} Let $\mathcal{C}={\eta _{0}}\mathcal{C}_{0}\oplus {\eta }_{1}%
\mathcal{C}_{1}\oplus \cdots \oplus {\eta _{s-1}}\mathcal{C}_{s-1}$ be an $m$%
-adic residue code of length $p$ over $\mathcal{R}_{q,s}.$ Then, $\mathcal{C}%
^{\bot }\subseteq \mathcal{C}$ if and only if $\mathcal{C}_{i}^{\bot
}\subseteq \mathcal{C}_{i}$ for each $i$.
\end{proposition}

\begin{proof}
It is clear that if, for each $i,$ $\mathcal{C}_{i}^{\bot }\subseteq
\mathcal{C}_{i},$ then $\mathcal{C}^{\bot }\subseteq \mathcal{C}.$ Suppose
that $\mathcal{C}^{\bot }\subseteq \mathcal{C}.$ Then, for each $0\leqslant
i\leqslant s-1,$ we get
\begin{eqnarray*}
&&\mathcal{C}{^{\bot }}\bmod\left( {{\eta _{0}},{\eta _{1}},\ldots ,{\eta
_{i-1}},{\eta _{i+1}},\ldots ,{\eta _{s-1}}}\right) \\
&\subseteq &\mathcal{C}\bmod\left( {{\eta _{0}},{\eta _{1}},\ldots ,{\eta
_{i-1}},{\eta _{i+1}},\ldots ,{\eta _{s-1}}}\right)
\end{eqnarray*}%
It follows that ${\eta _{i}}\mathcal{C}_{i}^{\bot }\subseteq {\eta _{i}}%
\mathcal{C}_{i}$ and $\mathcal{C}_{i}^{\bot }\subseteq \mathcal{C}_{i}.$
\end{proof}

Recall that for a cyclic code $\mathcal{C}$ over $\mathbb{F}_{q}$ with
defining set $Z,$ $\mathcal{C}^{\bot }\subseteq \mathcal{C}$ if and only if $%
-Z\cap Z=\emptyset $ \cite{ketkar}. Then, for an odd-like class-$I$ $m$-adic
residue code with defining set $Z=Q_{i}$, the condition to contain its dual
reduces $-Q_{i}\cap Q_{i}=\emptyset ,$ which is only possible when $-1\notin
{Q_{i}}$. The following explores when the case $-1\notin {Q_{i}}$ is
satisfied.

\begin{proposition}
\label{prop2} Let $p$ be a prime and $b$ be a primitive element of $\mathbb{Z%
}_{p}^{\ast }.$ Let $Q_{0}$ be the set of nonzero $m$-adic residues modulo $%
p $ with $m\geqslant 2$, $m\in \mathbb{Z}$ and $m|(p-1)$. Let $%
Q_{i}=b^{i}Q_{0} $. In this case, if $m$ does not divide ${{\left( {p-1}%
\right) }%
\mathord{\left/{\vphantom
{{\left( {p - 1} \right)} 2}} \right.\kern-\nulldelimiterspace}2}$, then $-{%
Q_{0}}\neq {Q_{0}}$ and so $-{Q_{i}}\neq {Q_{i}}$ for each $0\leqslant
i\leqslant m-1.$
\end{proposition}

\begin{proof}
Suppose that $m\nmid{{\left( {p-1}\right) }%
\mathord{\left/{\vphantom
{{\left( {p - 1} \right)} 2}} \right.\kern-\nulldelimiterspace}2}$ and $-{%
Q_{0}}={Q_{0}}.$ Then, clearly, $-1\notin {Q_{0}}$. Since ${b^{{{\left( {p-1}%
\right) }%
\mathord{\left/{\vphantom {{\left( {p - 1} \right)} 2}}
\right.\kern-\nulldelimiterspace}2}}}=-1$, for some $1\leqslant j\leqslant
p-2$, we get ${b^{jm}}={b^{{{\left( {p-1}\right) }%
\mathord{\left/{\vphantom
{{\left( {p - 1} \right)} 2}} \right.\kern-\nulldelimiterspace}2}}}.$ This
gives the congruence $jm-{{\left( {p-1}\right) }%
\mathord{\left/{\vphantom
{{\left( {p - 1} \right)} 2}} \right.\kern-\nulldelimiterspace}2}\equiv
0\left( {\bmod\left( {p-1}\right) }\right) ,$ where $m|p-1$ and $1\leqslant
j\leqslant p-2$. However, this is a contradiction because the last
congruence has no solution since $m|(p-1)$ and $m\nmid{{\left( {p - 1}
\right)}
\mathord{\left/{\vphantom {{\left( {p - 1} \right)} 2}}
\right.\kern-\nulldelimiterspace} 2}$. To complete the proof, suppose that $-%
{Q_{0}}\neq {Q_{0}}$ but $-{Q_{i}}={Q_{i}}$. Then, $-{b^{i}}\in {Q_{i}}$
implies that $-1\in {Q_{0}}$, which is a contradiction.
\end{proof}

As a result of Proposition \ref{prop2}, we rearrange Proposition \ref{prop1}
as following:

\begin{proposition}
\label{prop3} Let $\mathcal{C}={\eta _{0}}\mathcal{C}_{0}\oplus {\eta }_{1}%
\mathcal{C}_{1}\oplus \cdots \oplus {\eta _{s-1}}\mathcal{C}_{s-1}$ be an
odd-like class-$I$ $m$-adic residue code of length $p$ over $\mathcal{R}%
_{q,s},$ where $\mathcal{C}_{i}$ is an odd-like class-$I$ $m$-adic residue
code of length $p$ over $\mathbb{F}_{q}$ with the defining set $Q_{i}$.
Then, $\mathcal{C}^{\bot }\subseteq \mathcal{C}$ if and only if $-{Q_{i}}%
\neq {Q_{i}}$ for each $0\leqslant i\leqslant s-1$.
\end{proposition}

To get quantum error correcting codes by using CSS code construction, we
need to carry the codes over the ring $\mathcal{R}_{q,s}$ onto the finite
field $\mathbb{F}_{q}$ via a Gray map preserving the orthogonality from $%
\mathcal{R}_{q,s}^{n}$ to $\mathbb{F}_{q}^{sn}.$ We here state such a Gray
map defined in \cite{goyal} and recall its some properties.

\begin{definition}
\cite{goyal} Let $\zeta $ be $\left( {s-1}\right) ^{th}$ primitive root of
unity over $\mathbb{F}_{q},$ where $\left( {s-1}\right) |\left( {q-1}\right)
.$ A Gray map $\Phi :\mathcal{R}{_{q,s}}\rightarrow \mathbb{F}_{q}^{s}$ is
defined by
\begin{equation*}
{a_{0}}+{a_{1}}v+\cdots +{a_{s-1}}{v^{s-1}}\rightarrow \left( {{a_{0}},{a_{1}%
},\ldots ,{a_{s-1}}}\right) MV,
\end{equation*}%
where $V$ is any $s\times s$ nonsingular matrix over $\mathbb{F}_{q}$ and $M$
is $s\times s$ nonsingular matrix
\begin{equation*}
\left( {%
\begin{array}{cccccc}
1 & 1 & 1 & 1 & \cdots  & 1 \\
0 & 1 & \zeta  & {{\zeta ^{2}}} & \cdots  & {{\zeta ^{s-2}}} \\
0 & 1 & {{\zeta ^{2}}} & {{{\left( {{\zeta ^{2}}}\right) }^{2}}} & \cdots  &
{{{\left( {{\zeta ^{2}}}\right) }^{s-2}}} \\
0 & 1 & {{\zeta ^{3}}} & {{{\left( {{\zeta ^{3}}}\right) }^{2}}} & \cdots  &
{{{\left( {{\zeta ^{3}}}\right) }^{s-2}}} \\
\vdots  & \vdots  & \vdots  & \vdots  & \cdots  & \vdots  \\
0 & 1 & {{\zeta ^{s-2}}} & {{{\left( {{\zeta ^{s-2}}}\right) }^{2}}} &
\cdots  & {{{\left( {{\zeta ^{s-2}}}\right) }^{s-2}}} \\
0 & 1 & 1 & 1 & 1 & 1%
\end{array}%
}\right) .
\end{equation*}

The Gray map from $\mathcal{R}_{q,s}^{n}$ to $\mathbb{F}_{q}^{sn}$ is the
componentwise extension of the Gray map $\Phi $.
\end{definition}

For an element $r\in \mathcal{R}_{q,s},$ its Gray weight ${w_{G}}\left(
r\right) $ is the Hamming weight of $\Phi \left( r\right) $. For a vector $%
c=\left( {{c_{0}},{c_{1}},\ldots ,{c_{n-1}}}\right) $ in $\mathcal{R}%
_{q,s}^{n},$ its Gray weight ${w_{G}}\left( c\right) $ is the sum of the
Gray weights of its components. The Gray distance ${d_{G}}\left( {c,d}%
\right) $ between two vectors $c$ and $d$ in $\mathcal{R}_{q,s}^{n}$ is the
Gray weight of their difference $c-d$ \cite{goyal}. The Gray distance ${d_{G}%
}\left( \mathcal{C}\right) $ of a code $\mathcal{C}$ is the minimum Gray
distance between distinct nonzero codewords in the code $\mathcal{C}.$ We
also have the following:

\begin{proposition}
\label{prop4}\cite{goyal} The Gray map $\Phi $ is an $\mathbb{F}_{q}$-linear
and distance preserving map from $\mathcal{R}_{q,s}^{n}$ to $\mathbb{F}%
_{q}^{sn}$. Moreover, if $V{V^{T}}=\lambda {I_{n}},$ where $\lambda \in {%
F_{q}}$, then $\Phi $ is a preserving orthogonality map from $\mathcal{R}%
_{q,s}^{n}$ to $\mathbb{F}_{q}^{sn}$.
\end{proposition}

Combining Theorem \ref{thr0}, Proposition \ref{prop3} and Proposition \ref%
{prop4}, we derive a class of quantum error correcting codes by applying CSS
codes construction after taking the $\Phi $-Gray images of class-$I$
odd-like $m$-adic residue codes over the ring $\mathcal{R}_{q,s}$ that
contain their duals.

\begin{theorem}
\label{thr5} Let $\mathcal{C}={\eta _{0}}\mathcal{C}_{0}\oplus {\eta }_{1}%
\mathcal{C}_{1}\oplus \cdots \oplus {\eta _{s-1}}\mathcal{C}_{s-1}$ be an
odd-like class-$I$ $m$-adic residue code of length $p$ over $\mathcal{R}%
_{q,s},$ where $\mathcal{C}_{i}$ is an odd-like class-$I$ $m$-adic residue
code of length $p$ over $\mathbb{F}_{q}$ with the defining set $Q_{i}$ and
the generator polynomial ${g_{i}}\left( x\right) .$ Then, if $-{Q_{i}}\neq {%
Q_{i}}$ for each $0\leqslant i\leqslant s-1$, then there exists a quantum
error correcting code with the parameters ${\left[ \kern-0.15em\left[ {%
ps,ps-2k,{d_{G}}}\right] \kern-0.15em\right] _{q}}$ where $%
k=\sum\limits_{i=0}^{s-1}{\deg {g_{i}}\left( x\right) }$ and $d_{G}$ is the
Gray distance of the code $\mathcal{C}.$
\end{theorem}

We are going to present some examples of quantum error correcting codes to
illustrate what we find in this chapter.

\begin{example}
Let $\mathcal{C}=\left( {1-{v^{2}}}\right) \mathcal{C}{_{0}}\oplus \left( {%
2v+2{v^{2}}}\right) \mathcal{C}{_{1}}\oplus \left( {v+2{v^{2}}}\right)
\mathcal{C}{_{2}}$ be an odd-like class-$I$ $4$-adic residue code of length $%
13$ over the ring $\mathcal{R}_{3,3}$, where $\mathcal{C}_{0}$, $\mathcal{C}%
_{1}$ and $\mathcal{C}_{2}$ are odd-like class-$I$ $4$-adic residue codes of
length $13$ over $\mathbb{F}_{3}$ with the generator polynomials ${g_{0}}%
\left( x\right) =2+x+{x^{2}}+{x^{3}}$, ${g_{1}}\left( x\right) =2+{x^{2}}+{%
x^{3}}$ and ${g_{2}}\left( x\right) =2+2x+{x^{3}}$, respectively. In this
case, since the defining sets of $\mathcal{C}_{0}$, $\mathcal{C}_{1}$ and $%
\mathcal{C}_{2}$ are ${Z_{0}}={Q_{0}}=\left\{ {1,3,9}\right\} $, ${Z_{1}}=2{%
Q_{0}}=\left\{ {2,6,5}\right\} $ and ${Z_{2}}=8{Q_{0}}=\left\{ {8,11,7}%
\right\} $, respectively, it follows from ${Z_{i}}\neq -{Z_{i}}$ and so $%
\mathcal{C}_{i}^{\bot }\subseteq \mathcal{C}{_{i}}$ for $0\leqslant
i\leqslant 2$ that $\mathcal{C}{^{\bot }}\subseteq \mathcal{C}.$ Taking
image of $\mathcal{C}$ under the Gray map $\Phi :\mathcal{R}{_{3,3}}%
\rightarrow {\mathbb{F}_{3}^{3}},$ ${r_{0}}+{r_{1}}v+{r_{2}}{v^{2}}%
\rightarrow \left( {{r_{0}},{r_{0}}+{r_{1}}+{r_{2}},{r_{0}}+2{r_{1}}+{r_{2}}}%
\right) $ and using Theorem \ref{thr5}, we get a ${\left[ \kern-0.15em\left[
{39,21,3}\right] \kern-0.15em\right] _{3}}$ quantum code.
\end{example}

\begin{example}
Let $\mathcal{C}=\left( {1-{v}}\right) \mathcal{C}{_{0}}\oplus {v\mathcal{C}%
_{1}}$ be an odd-like class-$I$ $6$-adic residue code of length $19$ over the
ring $\mathcal{R}_{7,2}$, where $\mathcal{C}_{0}$ and $\mathcal{C}_{1}$ are
odd-like class-$I$ $6$-adic residue codes of length $19$ over $\mathbb{F}_{7}
$ with the generator polynomials ${g_{0}}\left( x\right) =6+4x+4{x^{2}}+{%
x^{3}}$ and ${g_{1}}\left( x\right) =6+x+4{x^{2}}+{x^{3}}$, respectively. In
this case, since the defining sets of $\mathcal{C}_{0}$ and $\mathcal{C}_{1}$
are ${Z_{0}}={Q_{0}}=\left\{ {1,7,11}\right\} $ and ${Z_{1}}=16{Q_{0}}%
=\left\{ {16,17,5}\right\} $, respectively, it follows from ${Z_{i}}\neq -{%
Z_{i}}$ and so $\mathcal{C}_{i}^{\bot }\subseteq \mathcal{C}{_{i}}$ for $%
0\leqslant i\leqslant 1$ that $\mathcal{C}{^{\bot }}\subseteq \mathcal{C}.$
Taking image of $\mathcal{C}$ under the Gray map $\Phi :\mathcal{R}{_{7,2}}%
\rightarrow {\mathbb{F}_{7}^{2}},$ ${r_{0}}+{r_{1}}v\rightarrow \left( {{%
r_{0}},{r_{0}}+{r_{1}}}\right) $ and using Theorem \ref{thr5}, we get a ${%
\left[ \kern-0.15em\left[ {38,26,3}\right] \kern-0.15em\right] _{7}}$
quantum code.
\end{example}

\begin{example}
Let $\mathcal{C}=\left( {1-{v^{2}}}\right) \mathcal{C}{_{0}}\oplus \left( {%
4v+4{v^{2}}}\right) \mathcal{C}{_{1}}\oplus \left( {3v+4{v^{2}}}\right)
\mathcal{C}{_{2}}$ be an odd-like class-$I$ $6$-adic residue code of length $%
19$ over the ring $\mathcal{R}_{7,3}$, where $\mathcal{C}_{0}$, $\mathcal{C}%
_{1}$ and $\mathcal{C}_{2}$ are odd-like class-$I$ $6$-adic residue codes of
length $19$ over $\mathbb{F}_{7}$ with the generator polynomials ${g_{0}}%
\left( x\right) =6+3x+6{x^{2}}+{x^{3}}$, ${g_{1}}\left( x\right) =6+5{x^{2}}+%
{x^{3}}$ and ${g_{2}}\left( x\right) =6+3x+3{x^{2}}{x^{3}}$, respectively.
In this case, since the defining sets of $\mathcal{C}_{0}$, $\mathcal{C}_{1}$
and $\mathcal{C}_{2}$ are ${Z_{0}}=2{Q_{0}}=\left\{ {2,14,3}\right\} $, ${%
Z_{1}}=4{Q_{0}}=\left\{ {4,9,6}\right\} $ and ${Z_{2}}=8{Q_{0}}=\left\{ {%
8,18,12}\right\} $, respectively, it follows from ${Z_{i}}\neq -{Z_{i}}$ and
so $\mathcal{C}_{i}^{\bot }\subseteq \mathcal{C}{_{i}}$ for $0\leqslant
i\leqslant 2$ that $\mathcal{C}{^{\bot }}\subseteq \mathcal{C}.$ Taking
image of $\mathcal{C}$ under the Gray map $\Phi :\mathcal{R}{_{7,3}}%
\rightarrow {\mathbb{F}_{7}^{3}},$ ${r_{0}}+{r_{1}}v+{r_{2}}{v^{2}}%
\rightarrow \left( {{r_{0}},{r_{0}}+{r_{1}}+{r_{2}},{r_{0}}+6{r_{1}}+{r_{2}}}%
\right) $ and using Theorem \ref{thr5}, we get a ${\left[ \kern-0.15em\left[
{57,39,3}\right] \kern-0.15em\right] _{7}}$ quantum code.
\end{example}

\begin{example}
Let $\mathcal{C}=\left( {1-{v}}\right) \mathcal{C}{_{0}}\oplus {v\mathcal{C}%
_{1}}$ be an odd-like class-$I$ $4$-adic residue code of length $19$ over the
ring $\mathcal{R}_{7,2}$, where $\mathcal{C}_{0}$ and $\mathcal{C}_{1}$ are
odd-like class-$I$ $4$-adic residue codes of length $29$ over $\mathbb{F}_{7}
$ with the generator polynomials ${g_{0}}\left( x\right) =6+6x+2{x^{2}}+4{%
x^{3}}+6{x^{6}}+x^{7}$ and ${g_{1}}\left( x\right) =6+5x+2{x^{2}}+2{x^{4}}%
+x^{5}+6{x^{6}}+x^{7}$, respectively. In this case, since the defining sets
of $\mathcal{C}_{0}$ and $\mathcal{C}_{1}$ are ${Z_{0}}={Q_{0}}=\left\{ {%
1,7,16,20,23,24,25}\right\} $ and ${Z_{1}}=2{Q_{0}}=\left\{ {%
2,3,11,14,17,19,21}\right\} $, respectively, it follows from ${Z_{i}}\neq -{%
Z_{i}}$ and so $\mathcal{C}_{i}^{\bot }\subseteq \mathcal{C}{_{i}}$ for $%
0\leqslant i\leqslant 1$ that $\mathcal{C}{^{\bot }}\subseteq \mathcal{C}.$
Taking image of $\mathcal{C}$ under the Gray map $\Phi :\mathcal{R}{_{7,2}}%
\rightarrow {\mathbb{F}_{7}^{2}},$ ${r_{0}}+{r_{1}}v\rightarrow \left( {{%
r_{0}},{r_{0}}+{r_{1}}}\right) $ and using Theorem \ref{thr5}, we get a ${%
\left[ \kern-0.15em\left[ {58,30,6}\right] \kern-0.15em\right] _{7}}$
quantum code.
\end{example}

\begin{example}
Let $\mathcal{C}=\left( {1-{v^{2}}}\right) \mathcal{C}{_{0}}\oplus \left( {%
4v+4{v^{2}}}\right) \mathcal{C}{_{1}}\oplus \left( {3v+4{v^{2}}}\right)
\mathcal{C}{_{2}}$ be an odd-like class-$I$ $4$-adic residue code of length $%
29$ over the ring $\mathcal{R}_{7,3},$ where $\mathcal{C}_{0},$ $\mathcal{C}%
_{1}$ and $\mathcal{C}_{2}$ are odd-like class-$I$ $4$-adic residue codes of
length $29$ over $\mathbb{F}_{7}$ with the generator polynomials ${g_{0}}%
\left( x\right) =6+5x+2x^{2}+2x^{4}+x^{5}+6x^{6}+x^{7}$, ${g_{1}}\left(
x\right) =6+x+3x^{4}+5x^{5}+x^{6}+x^{7}$ and ${g_{2}}\left( x\right)
=6+x+6x^{2}+5x^{3}+5x^{5}+2x^{6}+x^{7}$, respectively. In this case, since
the defining sets of $\mathcal{C}_{0}$, $\mathcal{C}_{1}$ and $\mathcal{C}%
_{2}$ are ${Z_{0}}=2{Q_{0}}=\left\{ {2,3,11,14,17,19,21}\right\} ,$ ${Z_{1}}%
=4{Q_{0}}=\left\{ {4,5,6,9,13,22,28}\right\} $ and ${Z_{2}}=8{Q_{0}}=\left\{
{8,10,12,15,18,26,27}\right\} $, respectively, it follows from ${Z_{i}}\neq -%
{Z_{i}}$ and so $\mathcal{C}_{i}^{\bot }\subseteq \mathcal{C}{_{i}}$ for $%
0\leqslant i\leqslant 2$ that $\mathcal{C}{^{\bot }}\subseteq \mathcal{C}.$
Taking image of $\mathcal{C}$ under the Gray map $\Phi :\mathcal{R}{_{7,3}}%
\rightarrow {\mathbb{F}_{7}^{3}},$ ${r_{0}}+{r_{1}}v+{r_{2}}{v^{2}}%
\rightarrow \left( {{r_{0}},{r_{0}}+{r_{1}}+{r_{2}},{r_{0}}+6{r_{1}}+{r_{2}}}%
\right) $ and using Theorem \ref{thr5}, we get a ${\left[ \kern-0.15em\left[
{87,45,6}\right] \kern-0.15em\right] _{7}}$ quantum code.
\end{example}

\section{Conclusion and Future Remarks}

In this paper, we generalize results of \cite{kuruz} to the non-chain
quotient ring $\mathcal{R}_{q,s},$ where $q$ is a prime power and $s$ is an
integer. We study the algebraic structure of $m$-adic residue codes over $%
\mathcal{R}_{q,s}$ in spite of the fact that previous papers worked over
finite fields. Further, these previous papers have more restrictions on the
parameters of the codes they obtained. We achieve to obtain some optimal
codes with respect to Griesmer bound for rings (see \cite{shiro}), which
makes this study even more remarkable. We seek a condition for the existence
of dual-containing $m$-adic residue codes over $\mathcal{R}_{q,s},$ and by
taking their Gray images, we construct a family of quantum error correcting
codes. By giving some examples, we complete the paper.

It can be shown that for some special cases, the generators of the $m$-adic
residue codes over $\mathcal{R}_{q,s}$ are palindromic. Thus, by help of
this fact DNA codes can be constructed as a future work.



\begin{thebibliography}{99}
\bibitem{ashraf} M. Ashraf, G. Mohamed, Quantum codes from cyclic codes over
${\mathbb{F}_3} + v{\mathbb{F}_3}$, Int. J. Quantum Inf., \textbf{12}(6),
1450042, (2014).

\bibitem{bag} T. Bag, A. K. Upadhyay, M. Ashraf, G. Mohammad, Quantum codes
from cyclic codes over the ring ${{{F_{p}}\left[ u\right] }%
\mathord{\left/
 {\vphantom {{{F_p}\left[ u \right]} {\left\langle {{u^3} - u} \right\rangle }}} \right.\kern-\nulldelimiterspace} {\left\langle {{u^3} - u} \right\rangle }%
}$, Asian-Eur. J. Math., \textbf{12}(07), 2050008, (2019).

\bibitem{A4} R. Brualdi, V. Pless, Polyadic codes, Discrete Appl. Math.,
Elsevier, \textbf{25} 3--71, (1989).

\bibitem{Chen} B. Chen, H.Q. Dinh, Y. Fan and S. Ling, Polyadic constacyclic
codes, IEEE Trans. Inf. Theory, Springer, \textbf{61} 4895--4904, (2015).

\bibitem{A6} X. Dong, L. Wenjie, Z. Yan, Generating idempotents of cubic and
quartic residue codes over field $\mathbb{F}_2$, Designs, Computer
Engineering and Applications, North China Computing Technology Institute,
\textbf{49} 41--44, (2013).

\bibitem{goyal} M. Goyal, M. Raka, Duadic codes over the ring $\mathbb{F}%
_{q}[u]/(u^{m}-u)$ and their gray images, Journal of Computer and
Communications, \textbf{4}(12), 50--62, (2016).

\bibitem{goyalquadratic} M. Goyal, M. Raka, Quadratic residue codes over the
ring $\mathbb{F}_{p}[u]/(u^{m}-u)$ and their Gray images, Cryptogr. Commun.,
Springer, \textbf{10} 343-355, (2018).

\bibitem{A1} V.R. Job, m-adic residue codes, IEEE Trans. Inf. Theory,
\textbf{38} 496--501, (1992).

\bibitem{kai} X. Kai, S. Zhu, Quaternary construction of quantum codes from
cyclic codes over ${\mathbb{F}_{4}}+u{\mathbb{F}_{4}}$, Int. J. Quantum
Inf., \textbf{9}(02), 689--700, (2011).

\bibitem{ketkar} A. Ketkar, A. Klappenecker, S. Kumar and P. K. Sarvepalli,
Nonbinary stabilizer codes over finite fields, IEEE Trans. Inform.
Theory, \textbf{52}(11), 4892--4914, (2006).

\bibitem{kuruz} F. Kuruz, E.S. Oztas and I. Siap, $m$-adic codes over $%
\mathbb{F}_q[v]/(v^2-v)$ and DNA codes, Bull. Korean Math. Soc., \textbf{37}
37--38, (2018).

\bibitem{B1} S. Ling and C. Xing, Coding theory: A first course, Cambridge
University Press, (2004).

\bibitem{Ling} S. Ling and C. Xing, Polyadic codes revisited, IEEE Trans.
Inf. Theory, Springer, \textbf{50} 200--207, (2004).

\bibitem{MacWilliams} F. J. MacWilliams and N.J.A. Sloane, The theory of
error-correcting codes, Elsevier, (1977).

\bibitem{A3} F. J. MacWilliams, Generalized quadratic residue codes, IEEE
Trans. Inf. Theory, \textbf{24} 730--737,(1978).

\bibitem{A2} V. Pless, Polyadic Codes, Algebraic Combinatorial Theory,
107--115, (1988).

\bibitem{qian1} J. Qian, Quantum codes from cyclic codes over ${\mathbb{F}_2}
+ v{\mathbb{F}_2}$, Journal of Information Computational Science, \textbf{10}%
(6), 1715--1722, (2013).


\bibitem{G1} M. Raka, L. Kathuria, M. Goyal, $(1-2u^{3})$-constacyclic codes
and quadratic residue codes over $\mathbb{F}_{p}[u]/(u^{4}-u)$, Cryptogr.
Commun., Springer, \textbf{9} 459--473, (2017).

\bibitem{sari2} M. Sar{\i}, I. Siap, On quantum codes from cyclic codes over
a class of nonchain rings, Bull. Korean Math. Soc., \textbf{53}(6),
1617--1628, (2016).

\bibitem{sari1} M. Sar{\i}, I. Siap, Quantum codes over a class of finite
chain rings, Quantum Information and Computation, \textbf{16}(1-2),
0039--0049, (2016).

\bibitem{shiro} K. Shiromoto, L. Storme, A Griesmer bound for linear codes
over finite quasi-Frobenius rings, Discrete Appl. Math., Elsevier, \textbf{%
128.1} 263--274, (2003).

\bibitem{tang} Y. Tang, T. Yao, S. Zhu, X. Kai, A family of constacyclic
codes over ${\mathbb{F}_{{2^m}}} + u{\mathbb{F}_{{2^m}}}$ and its
application to quantum codes, Chinese Journal of Electronics, \textbf{29}%
(1), 114--120, (2020).

\bibitem{A5} A.J. van Zanten, A. Bojilov, S.M. Dodunekov, Generalized
residue and t-residue codes and their idempotent generators, Des. Codes
Cryptogr., Springer, \textbf{75} 315--334, (2015).


\bibitem{xunru} Y. Xunru, M. Wenping, Gray map and quantum codes over the
ring ${\mathbb{F}_2} + u{\mathbb{F}_2} + {u^2}{\mathbb{F}_2}$, International
Joint Conference of IEEE TrustCom, pp. 897--899, (2011).



\end{thebibliography}
\end{document}